\documentclass[aps,prb,showpacs,superscriptaddress,twocolumn]{revtex4}
\usepackage{amsfonts}
\usepackage{txfonts}
\usepackage{amssymb}
\usepackage{amsbsy} 
\usepackage{epsfig}

\def\be{\begin{equation}} \def\ee{\end{equation}}
\def\bea{\begin{eqnarray}} \def\eea{\end{eqnarray}}

\def\nn{\nonumber}

\def\bk{{\bf k}}
\def\br{{\bf r}}
\def\bx{{\bf x}}

\def\la{\langle}
\def\ra{\rangle}

\def\rw{\rightarrow}

\begin{document}

\title{Topological Invariants and Ground-State Wave Functions of Topological Insulators on a Torus}

\author{Zhong Wang}
\altaffiliation{wangzhongemail@gmail.com} \affiliation{ Institute for
Advanced Study, Tsinghua University, Beijing,  China, 100084}

\author{Shou-Cheng Zhang$^{1,}$}

\affiliation{
Department of Physics, Stanford University, Stanford, CA 94305}

\date{ \today}

\begin{abstract}

We define topological invariants in terms of the ground states
wavefunctions on a torus. This approach leads to precisely defined
formulas for the Hall conductance in four dimensions and the
topological magneto-electric $\theta$ term in three dimensions, and
their generalizations in higher dimensions. They are valid in the
presence of arbitrary many-body interaction and disorder. These
topological invariants systematically generalize the two-dimensional
Niu-Thouless-Wu formula, and will be useful in numerical calculations
of disordered topological insulators and strongly correlated
topological insulators, especially fractional topological insulators.

\end{abstract}


\maketitle

\section{Introduction}\label{sec:intro}

Topological insulators are among the major recent developments in
condensed matter physics\cite{qi2010a,hasan2010,qi2011}. The physics
of topological insulators started with noninteracting
systems\cite{kane2005a,bernevig2006a,kane2005b,bernevig2006c,konig2007,fu2006,moore2007,qi2008,fu2007b,roy2009a,schnyder2008,kitaev2009},
for which simple and calculable topological invariants have been
invaluable tools. More recently, it became clear that the interplay
between topology and many-body interaction is a still richer
field\cite{raghu2008,shitade2009,zhang2009b,seradjeh2009,
pesin2010,fidkowski2010,hohenadler2011,wang2010b,sun2009,
li2010,dzero2010,rachel2010,regnault2011,zhang2011,guo2011,
guo2012,sheng2011,zheng2011,yu2011,levin2009,
maciejko2010a,swingle2011, varney2011, griset2012,
neupert2012,go2012,chen2013,chen2013a,wang2011b,budich2012,
wang2012b,wang2011d,hohenadler2012,kargarian2011,
assaad2013,budich2013a,hung2013,hung2013a,werner2013,lu2013,lang2013,
hohenadler2013,zhang2013,wang2012f,araujo2013,zhu2013,xu2013,xu2013a,oon2012,bi2013},
therefore, it is highly desirable to develop topological invariants
that are valid in the presence of strong interaction.

The root state of three-(spatial)-dimensional (3D) and
two-(spatial)-dimensional (2D) topological insulators with time
reversal symmetry is the four-(spatial)-dimensional (4D) quantum Hall
(QH) state\cite{zhang2001,qi2008} from which the topological field
theory of 3D and 2D insulators can be obtained by the procedure of
``dimensional reduction''\cite{qi2008}. The electromagnetic effective
action of the 4D QH effect reads\cite{bernevig2002,qi2011} \bea
S_{{\rm eff}}=\frac{\sigma_{4D}}{24\pi^2}\int
dtd^4x\epsilon^{\mu\nu\rho\sigma\tau}A_\mu\partial_\nu A_\rho
\partial_\sigma A_\tau \label{CS} \eea where we have adopted the
units that the electric charge $e$, the Planck constant $h$, and the
light velocity $c$ are all unity. The coefficient $\sigma_{4D}$ is
referred to as the ``4D Hall conductance''(or the 4D Hall
coefficient). Physically, the 4D QH effect has the nonlinear
topological electromagnetic response\cite{qi2008} $j^\mu=\frac{\delta
S_{{\rm eff}} }{\delta A_\mu}
=\frac{\sigma_{4D}}{8\pi^2}\epsilon^{\mu\nu\rho\sigma\tau}\partial_\nu
A_\rho\partial_\sigma A_\tau$ in the bulk, which is described
naturally by the Chern-Simons effective action. If a nontrivial 4D QH
insulator is cut open in one direction, there are $|\sigma_{4D}|$
copies of 3D chiral fermions (Weyl fermions) modes localized at the
boundary. These boundary modes are  close analogues of the 1D chiral
edge states\cite{halperin1982,wen1990} of 2D QH. In fact, the QH
effect can be generalized to all even spatial dimensions, whose
boundary modes are chiral fermions in odd spatial dimensions.

In the noninteracting limit, the explicit formula for $\sigma_{4D}$
has been obtained by Qi, Hughes and Zhang as\cite{qi2008} \bea
\sigma_{4D}=c_2\equiv \frac{1}{32\pi^2}\int d^4k \epsilon^{ijkl}{\rm
Tr}f_{ij}f_{kl} \label{noninteracting} \eea where $f_{ij}$ is the non-Abelian Berry curvature defined in terms of the noninteracting Bloch states. \footnote{The lower-case
letter $c_2$ refers to the noninteracting limit, which should not be confused with the upper-case $C_2$ defined in Sec.\ref{sec:more}.}

Now a natural question
arises: Can we find a formula for $ \sigma_{4D}$ that is precisely
defined in the presence of arbitrary interaction and disorder? Such a
formula, if exists, will be especially desirable for the
investigation of fractional quantum Hall states in 4D. More
importantly, it may also shed light on strongly interacting
topological insulators in lower dimensions.

The same question also arises for the 3D topological insulators,
whose effective topological responses theory is given by\cite{qi2008}
\bea S_{{\rm eff}}=\frac{1}{8\pi^2}\int dtd^3x
\epsilon^{ijkl}\theta\partial_i A_j\partial_k A_l
=\frac{1}{4\pi^2}\int dtd^3x \theta{\bf E}\cdot{\bf B} \label{theta}
\eea This topological effective action describes the quantized
topological magnetoelectric effect, in which an electric field
induces a magnetization with universal constant of
proportionality\cite{qi2008}.

In the noninteracting limit, $\theta$ has a simple
expression\cite{qi2008,essin2009,wang2010a} \bea \theta =
\frac{1}{4\pi } \int d^{3}k\epsilon^{ijk} \textrm{Tr}\{[\partial_i
a_{j}(k)+\frac{2}{3}i a_{i}(k)a_{j}(k))] a_{k}(k)\} \label{CS-3d}
\eea which is a 3D Chern-Simons term. In the presence of
time-reversal symmetry, this Chern-Simons term is quantized and has
been shown to be equivalent\cite{wang2010a} to the $Z_2$ topological
invariant\cite{fu2007b}.  The natural question is: Is there a formula
for $\theta$ that is valid in the presence of arbitrary interaction
and disorder? From the experimentalist's perspective, this question
is more urgent than the 4D QH case, because many 3D topological
insulators have been realized in experiments, and the
electron-electron interaction has been playing more important roles.

To partially answer these questions,  interacting topological
invariants expressed in terms of Green's function at zero-frequency
(namely the ``topological Hamiltonian''\cite{wang2013}) for
interacting insulators have been
proposed\cite{wang2012a,wang2012d,wang2012}, which provide an
efficient approach for topological invariants of various topological
insulators and superconductors [See, e.g.
Ref.\cite{go2012,budich2013a,werner2013,hung2013,hung2013a,lu2013,lang2013,hohenadler2013,nakosai2013,manmana2012,deng2013,yoshida2013,budich2013}
for applications]. However, there are several shortcomings of that
approach. First, it cannot be directly applied to disordered systems
in which the momentum $\bk$ in the single-particle Green's function
is not a good quantum number.\footnote{Enlarging the unit cell can
partially overcome this difficulty, which enables an extension of
this approach to disordered systems.} Second, it is unclear whether
or not that approach may fail for some fractional topological states.

In Ref.\cite{niu1985}, Niu, Thouless and Wu found for the 2D QH a
topological invariant (the first Chern number) expressed in terms of
ground state wavefunction under twisted boundary condition, which is valid in the presence of
arbitrary interaction and disorder. \footnote{Twisted boundary conditions have also found applications elsewhere, see e.g. Ref.\cite{kohn1964,hetenyi2013}. } To search for the general
formulas for $\sigma_{4D}$ in 4D and $\theta$ in 3D, a hopeful
approach is to generalize their formula to higher dimensions.
However, as we will see later, the most straightforward 4D
generalization of their formula, namely the generalization of the 2D
phase twisting $(\theta_1,\theta_2)$ to the 4D phase twisting
$(\theta_1,\theta_2,\theta_3,\theta_4)$ [see Eq.(\ref{standard})],
cannot produce the 4D Hall conductance $\sigma_{4D}$. Due to this
difficulty, it is unclear how this approach can be generalized to
higher-dimensional topological states.

In this paper we propose general topological invariants for higher
dimensional topological insulators in terms of ground state
wavefunctions. The boundary conditions adopted here are not the
standard one used in Ref.\cite{niu1985}, which is a pure gauge with
vanishing field strength.  Using these new boundary conditions [see
Sec.\ref{sec:4d} and Sec.\ref{sec:3d}], we obtain for $\sigma_{4D}$
and $\theta$ simple formulas expressed in terms of the ground state
wavefunction on a torus [see
Eq.(\ref{interacting}),Eq.(\ref{fractional-4d}),Eq.(\ref{3d}),
Eq.(\ref{3d-fractional}), etc]. We also generalize these formulas to
higher dimensions[see Eq.(\ref{d-d}), etc]. These topological
invariants are valid in the presence of arbitrary interaction and
disorder, thus they can be applied to topological states with strong
disorders and strongly correlated topological states including
fractionalized states. Unexpectedly, the generalized formula for
$\sigma_{4D}$ appears not as a second Chern number, but as the
\emph{difference} between two \emph{first} Chern numbers
[Eq.(\ref{interacting}),Eq.(\ref{fractional-4d})]. Similarly, the
formula for $\theta$ does not appear as a Chern-Simons form, but as
the difference between two winding numbers[Eq.(\ref{3d}),
Eq.(\ref{3d-fractional})].

The rest part of this paper is organized as follows. In
Sec.\ref{sec:4d} we study the 4D QH and define the topological
invariant for integer QH in 4D.  In Sec.\ref{sec:4d-dirac} we test
this topological invariant in two noninteracting models. We then
generalize the 4D topological invariant to higher dimensional QH
effects in Sec.\ref{sec:higher}. In Sec.\ref{sec:fractional} we
present the topological invariants for fractional quantum Hall
effects. A different boundary condition is investigated in
Sec.\ref{sec:more}, which leads to 4D topological invariant unrelated
to the 4D Hall conductance. The next two Sections, namely
Sec.\ref{sec:1d} and Sec.\ref{sec:3d}, is devoted to 1D and 3D
$\theta$ term respectively.

\section{4D Hall coefficients $\sigma_{4D}$ expressed in terms of the ground state wavefunction}\label{sec:4d}

In this section we describe the topological invariant defined in
terms of the ground state wave function of a 4D insulator on a torus
with generalized twisted boundary conditions. For simplicity, in this
section we assume that the ground sate is unique, while the cases
with ground state degeneracy will be studied in
Sec.\ref{sec:fractional}. We take the system to be a 4D torus with
circumference $L_1,L_2,L_3,L_4$ along the $x_1,x_2,x_3,x_4$ direction
respectively. We take the generalized twisted boundary condition
parameterized by $(\theta_1,\theta_2,\phi)$ as follows.
\footnote{Note that there are two ways to define the twisted boundary
condition. The first is to put the twisted phase factor
$(\theta_1,\theta_2,\phi)$ in the wavefunction, as we did in
Eq.(\ref{condition-1}), Eq.(\ref{condition-2}), and
Eq.(\ref{condition-3}). The second way is to add phase parameters in
the Hamiltonian instead of the wavefunction. These two ways are
equivalent and can be translated into each other. In fact, a gauge
transformation of the wavefunction changes the twisted boundary
condition to the periodic boundary condition, at the price of adding
the twisted phase factor to the Hamiltonian. } First, for
$i=1,2$,\bea \Psi(\br_1,\cdots,\br_k +L_i\hat{\bx}_i,\cdots,\br_N;
\theta_1,\theta_2,\phi) \nn \\ = \exp(i\theta_i) \Psi
(\br_1,\cdots,\br_k,\cdots,\br_N; \theta_1,\theta_2,\phi)
\label{condition-1} \eea where $\br_k$ is the coordinate of the
$k$-th particle ( other arguments such as spin are not shown here for
simplicity of notation ), $N$ is the total particle number, and
$\hat{\bx}_i$ is the unit vector along the $x_i$ direction. This
condition is the same as the one adopted in Ref.\cite{niu1985}.
Second, \bea \Psi (\br_1,\cdots,\br_k +L_3\hat{\bx}_3,\cdots,\br_N;
\theta_1,\theta_2,\phi) \nn \\ = \exp(-i\phi\frac{ x_4}{L_4}) \Psi
(\br_1,\cdots,\br_k,\cdots,\br_N; \theta_1,\theta_2,\phi)
\label{condition-2} \eea Since $x_4\equiv x_4+L_4$ on the torus,  the
flux $\phi$ has to be quantized as $n\phi_0$, where the unit flux
$\phi_0\equiv 2\pi$, and $n$ is an integer. Lastly \bea \Psi
(\br_1,\cdots,\br_k +L_4\hat{\bx}_4,\cdots,\br_N;
\theta_1,\theta_2,\phi) \nn \\ = \Psi
(\br_1,\cdots,\br_k,\cdots,\br_N; \theta_1,\theta_2,\phi)
\label{condition-3} \eea Physically, these twisted boundary
conditions tell us that there is a gauge potential $A_i
=\theta_i/L_i$ along the $x_i$($i=1,2$) direction, and a gauge
potential $A_3=-\phi\frac{ x_4}{L_3 L_4}$ along the $x_3$ direction,
in other words, there is a magnetic flux $\phi$ inside any  2D torus
$T_{34}$ whose coordinates are $(X_1,X_2,x_3,x_4)$ with fixed
$(X_1,X_2)$.

Before proceeding to our central results, let us briefly outline the
motivations of the boundary conditions given in
Eq.(\ref{condition-1}), Eq.(\ref{condition-2}), and
Eq.(\ref{condition-3}). The first motivation is that the
$(\theta_1,\theta_2,\theta_3,\theta_4)$ boundary condition [see
Sec.\ref{sec:more}] does not produce the 4D Hall conductance. The
second motivation is the intuitive relation between the 4D Hall
effect and the 2D Hall effect. In Eq.(\ref{CS}), if we take $A_3,
A_4$ to be independent on $x_0,x_1,x_2$, and at the same time take
$A_0,A_1,A_2$ to be independent on $x_3,x_4$, then there is a
``dimensional reduction'' \footnote{This is analogous to the
Kaluza-Klein compactification.} of the 4D Chern-Simons term to the 2D
Chern-Simons term:
$\sigma_{4D}\epsilon^{\mu\nu\rho\sigma\tau}A_\mu\partial_\nu A_\rho
\partial_\sigma A_\tau\rightarrow \sigma_{4D} B_{34}\epsilon^{\mu\nu\rho}A_\mu\partial_\nu
A_\rho$ (up to a numerical factor), where $B_{34}\equiv \partial_3
A_4 -\partial_4 A_3 $, and the indices $\mu,\nu,\rho$ in
``$\epsilon^{\mu\nu\rho}$'' take value $0,1,2$. According to this
argument, in our boundary conditions given in Eq.(\ref{condition-1}),
Eq.(\ref{condition-2}), and Eq.(\ref{condition-3}), we have taken
$\partial_3 A_4 -\partial_4 A_3 =\phi/L_3L_4$, thus we have the
dimensional reduction
$\sigma_{4D}\epsilon^{\mu\nu\rho\sigma\tau}A_\mu\partial_\nu A_\rho
\partial_\sigma A_\tau\rightarrow \sigma_{4D} \phi\epsilon^{\mu\nu\rho}A_\mu\partial_\nu
A_\rho$.  Intuitively, we have the evident identity \bea
\frac{\partial}{\partial \phi} ( \sigma_{4D}
\phi\epsilon^{\mu\nu\rho}A_\mu\partial_\nu A_\rho ) =
\sigma_{4D}\epsilon^{\mu\nu\rho}A_\mu\partial_\nu A_\rho \eea Since
the right hand side of this equation is a 2D Chern-Simons term, it
seems that we can calculate $\sigma_{4D}$ using well-known results of
2D quantum Hall effects.  In practice, however, it is impossible to
take the derivative with respect to $\phi$ because $\phi$ is
quantized, i.e. $\phi$ takes only discrete values. To resolve this
difficulty, we will take a difference instead of a derivative (see
below).

Now our task is to formulate these intuitive arguments as a precise
mathematical framework. We can define the Berry connection \bea
a_i(\theta_1,\theta_2,\phi) =-i\la\Psi( \theta_1,\theta_2,\phi)|
\partial_{\theta_i} |\Psi (\theta_1,\theta_2,\phi)\ra \label{connection} \eea and the Berry curvature \bea
F_{ij}(\theta_1,\theta_2,\phi)=\frac{\partial a_j}{\partial \theta_i}
-\frac{\partial a_i}{\partial \theta_j} \eea from which we can define
a first Chern number \bea C(\phi)=\frac{1}{2\pi} \int_0^{2\pi}
d\theta_1 d\theta_2 F_{12}(\theta_1,\theta_2,\phi) \label{chern-4d}
\eea where we have chosen the notation ``$C$'' instead of ``$C_1$''
to distinguish $C$ with the first Chern number appearing in the
2D quantum Hall effects\cite{niu1985}.

With these preparations, the general formula for $\sigma_{4D}$
appearing in Eq.(\ref{CS}) is proposed as \bea \sigma_{4D}=
C(\phi_0)-C(0) \label{interacting} \eea This is the difference
between two first Chern numbers, the first of which is the Chern
number with a unit flux $\phi_0\equiv 2\pi$ in $T_{34}$, and the
second is the Chern number without this flux, in other words,
Eq.(\ref{interacting}) measures the jump of the first Chern number
after inserting a flux $\phi_0$ in $T_{34}$. The necessity of the
second term $C(0)$ in Eq.(\ref{interacting}) can be easily
appreciated in a noninteracting model [see Eq.(\ref{4d-model-2})] to
be presented in Sec.\ref{sec:4d-dirac}. It is also useful to note
that $C(0)$ may be zero if the ground state has certain symmetries.
For instance, if there is time reversal symmetry, we have $C(0)=0$
and $\sigma_{4D}= C(\phi_0)$.

Eq.(\ref{interacting}) is expressed in terms of the Berry phase of
ground states wavefunctions on a torus, which is well-defined in the
presence of arbitrary interaction and disorder. \footnote{ We can
also write $\sigma_{4D} = [C(n\phi_0)-C(0)]/n$ for any integer $n$.
For simplicity, this will not be pursued in the present paper. }
Eq.(\ref{interacting}) can also be written equivalently as \bea
\sigma_{4D} = \frac{1}{2\pi} \int_0^{2\pi} d\theta_1 d\theta_2
[F_{12}(\theta_1,\theta_2,\phi_0) - F_{12}(\theta_1,\theta_2,0)]
\label{interacting-2} \eea Eq.(\ref{interacting}) and
Eq.(\ref{interacting-2}) are among the central equations of the
present paper.

Several remarks about Eq.(\ref{interacting}) are in order. The
noninteracting topological invariant for the 2D quantum Hall effect,
namely the TKNN invariant\cite{thouless1982}, is expressed as the
first Chern number in the Brillouin zone. The Niu-Thouless-Wu
formula\cite{niu1985}, as a generalization of the TKNN invariant, is
again a first Chern number. Given the second Chern number in
Eq.(\ref{noninteracting}) for the 4D noninteracting quantum Hall
effect, we may try to express the 4D Hall coefficient $\sigma_{4D}$
as a second Chern number on certain parameter space, for an
interacting system. However, this attempt turns out to be unfruitful.
Instead, the topological invariant defined in Eq.(\ref{interacting}),
which gives $\sigma_{4D}$, is the difference between two first Chern
numbers.

Let us conclude this section with a side remark that the Laughlin's
gauge argument\cite{laughlin1981} can also be generalized to 4D QH.
The boundary condition in the $x_3,x_4$ direction are the same as
given by Eq.(\ref{condition-2}) and Eq.(\ref{condition-3}), but the
system is open along the $x_2$ direction. When we do the adiabatic
evolution $\theta_1\rw \theta_1+2\pi$, the charge transferred from
the boundary $x_2=0$ to $x_2=L_2$ is denoted as $\Delta Q(\phi)$. The
Hall conductance is given as $\sigma_{4D}=\Delta Q(\phi_0)-\Delta
Q(0)$.

\section{The noninteracting limit: Two simple models}\label{sec:4d-dirac}

In this section we will check in two simple noninteracting models
[Eq.(\ref{4d-dirac}) and Eq.(\ref{4d-model-2})] that
Eq.(\ref{interacting}) gives the same result as
Eq.(\ref{noninteracting}), as it should do in the noninteracting
limit. Incorporating well-known results of topological classification of noninteracting insulators, we will show that Eq.(\ref{interacting}) reduces to Eq.(\ref{noninteracting}) for all noninteracting 4D insulators.

First let us consider a noninteracting Hamiltonian for 4D QH\cite{qi2008}
\bea h(\bk)=  v\sum_{i=1}^4 \sin k_i \Gamma^i  + M(\bk) \Gamma^0
\label{4d-dirac} \eea where $M(\bk)=m+4-\sum_{i=1}^4\cos k_i$,  $v$
and $m$ being parameters of the Hamiltonian, and $k_i\in [0,2\pi]$ is
the $i$-th momentum of the free Bloch state (the lattice constant has
been taken as unity). The Gamma matrices here satisfy the identities
$\{\Gamma^\mu,\Gamma^\nu\}=2\delta^{\mu\nu}$. For our convenience we
choose the representation $\Gamma^1=\tau^1, \Gamma^2=\tau^2,
\Gamma^3=\tau^3\sigma^1,
\Gamma^4=\tau^3\sigma^2,\Gamma^0=\tau^3\sigma^3$.

Instead of solving the model numerically in the real space, which is
less illuminating for our purpose, let us do calculation in the limit
that $|m|$ is significantly smaller than unity. In this limit we can
keep only the $\bk$-linear terms near $\bk=0$, and the Dirac
Hamiltonian reads \bea h(\bk)\approx v(k_1\tau^1+k_2\tau^2) +
\tau^3(vk_3\sigma^1+vk_4\sigma^2+m\sigma^3) \label{linear} \eea

In the presence of twisted boundary conditions, the momenta should be
replaced by $k_i \rw -iD_i = -i(\partial_i-A_i)$. Let us calculate
the first term $C(\phi_0)$ of Eq.(\ref{interacting}) for the Dirac
Hamiltonian in Eq.(\ref{linear}). In this linear-$\bk$ limit, we can
first solve the Hamiltonian $h'(k_3,k_4)=
vk_3\sigma^1+vk_4\sigma^2+m\sigma^3$, whose eigenvalues
read\cite{neto2009} \bea E_0=m; \,\,\, E_{n\pm} =
\pm\sqrt{m^2+2nBv^2}\, (n=1,2,\cdots) \label{eigen}\eea where
$B=\phi_0/L_3 L_4$. The corresponding eigen-wavefunctions are
$(\psi_0,0)^T$ and $(\psi_n,\pm\psi_{n-1})^T$, where $\psi_n$ is the
wavefunction of the $n$-th Landau level of Schrodinger
particles\cite{neto2009}, whose precise forms do not concern us for
our purpose. It is useful to note that when $m=0$, the existence of
the zero mode $E_0$ is guaranteed by the Atiyah-Singer index theorem.
Inputting the eigenvalues given in Eq.(\ref{eigen}) into the second
parenthesis in Eq.(\ref{linear}), we have a serial of 2D
Hamiltonians \bea h_0 &=&  v(k_1\tau^1 + k_2\tau^2)+m\tau^3; \nn\\
h_{n\pm}&=& v(k_1\tau^1 + k_2\tau^2)+E_{n\pm}\tau^3\,(n=1,2,\cdots )
\eea

The value of $C(\phi_0)$ can be obtained as the summation of the
first Chern number of $h_0$ and $h_{n\pm}$, namely $\frac{1}{2}[{\rm
sgn}(E_0)+\sum_{n }\sum_{\alpha=\pm} {\rm sgn}(E_{n\alpha})]
=\frac{1}{2}{\rm sgn}(m)$, thanks to the fact that the ground state
wavefunctions is a Slate determinant of Bloch states in the
noninteracting cases. In this calculation we have not been careful
about the high energy regularization, thus we can only assert that
$C(\phi_0)=\frac{1}{2}{\rm sgn}(m)+{\rm constant}$. Since we require
$C(\phi_0)=0$ as $m\rw +\infty$, we have $C(\phi_0)=\frac{1}{2}[{\rm
sgn}(m)-1]$. Similarly we can obtain that $C( 0)=0$, therefore we
have \bea \sigma_{4D}=C(\phi_0)-C(0)=\frac{1}{2}[{\rm sgn}(m)-1] \eea
which is the same as $c_2$ obtained\cite{qi2008} from
Eq.(\ref{noninteracting}) [see also Ref.\cite{li2012} for
calculations for a different model using charge pumping.]

Let us move to the second noninteracting model, which will explain
the reason why we must include the second term $C(0)$ in
Eq.(\ref{interacting}). The simple model has the free Hamiltonian
\bea h(\bk)=v(\sin k_1\tau^1+\sin k_2\tau^2) + (m+2-\cos k_1-\cos
k_2)\tau^3 \label{4d-model-2} \eea which is independent on $k_3$ and
$k_4$.  If we take $m=-0.1$, then it is obvious that both $C(\phi_0)$
and $C(0)$ are nonzero, however, they are equal, therefore
$\sigma_{4D}=C(\phi_0)-C(0)=0$. From Eq.(\ref{noninteracting}), it is
obvious that we have $\sigma_{4D}=c_2=0$, therefore,
Eq.(\ref{interacting}) and Eq.(\ref{noninteracting}) produce the same
result in this example.

Although we have only explicitly checked that Eq.(\ref{interacting})
reduces to Eq.(\ref{noninteracting}) in Dirac models, it is possible
to make a more general statement that Eq.(\ref{interacting}) is
always equivalent to Eq.(\ref{noninteracting}) in the noninteracting
limit. In fact, as has been shown in Ref.\cite{kitaev2009,ryu2010},
there is a Dirac-Hamiltonian representative in each class of the 4D
QH insulators, which means that any noninteracting Hamiltonian for 4D
insulator can always be smoothly connected to a Dirac Hamiltonian.
Therefore, equivalence between Eq.(\ref{interacting}) and
Eq.(\ref{noninteracting}) in Dirac model implies their equivalence
for all noninteracting Hamiltonians. In the presence of interaction, however, Eq.(\ref{noninteracting}) loses definition, while Eq.(\ref{interacting}) remains useful.

\section{Quantum Hall effect in $d=2l+2$ spatial
dimensions}\label{sec:higher}

Eq.(\ref{interacting}) can be generalized to $d=2l+2$ spatial
dimensions. The boundary conditions for the $(x_1,x_2)$ direction
given in Eq.(\ref{condition-1}) are unchanged, while the boundary
conditions for other directions are defined  as \bea
\Psi(\br_1,\cdots,\br_k +L_{2j+1}\hat{\bx}_{2j+1},\cdots,\br_N;
 \theta_1,\theta_2,\phi_1,\cdots,\phi_l ) \nn
\\ = \exp(-i\phi_j\frac{ x_{2j+2}}{L_{2j+2}})
\Psi (\br_1,\cdots,\br_k,\cdots,\br_N;
\theta_1,\theta_2,\phi_1,\cdots,\phi_l) \label{condition-2-d} \eea
and \bea \Psi (\br_1,\cdots,\br_k
+L_{2j+2}\hat{\bx}_{2j+2},\cdots,\br_N;
\theta_1,\theta_2,\phi_1,\cdots,\phi_l) \nn \\ = \Psi
(\br_1,\cdots,\br_k,\cdots,\br_N;
\theta_1,\theta_2,\phi_1,\cdots,\phi_l) \label{condition-3-d} \eea
for $j=1,2,\cdots l$. Physically, these conditions means that there
is a flux $\phi_j$ in the 2D torus $T_{2j+1,2j+2}$. We can define the
Berry connection \bea && a_i(\theta_1,\theta_2,\phi_1,\cdots,\phi_l)
\nn
\\ && =-i\la\Psi( \theta_1,\theta_2,\phi_1,\cdots,\phi_l)|
\partial_{\theta_i} |\Psi( \theta_1,\theta_2,\phi_1,\cdots,\phi_l)\ra \eea
for $i=1,2$, and a first Chern number \bea
C(\phi_1,\cdots,\phi_l)=\frac{1}{2\pi} \int_0^{2\pi} d\theta_1
d\theta_2 F_{12}(\theta_1,\theta_2,\phi_1,\cdots,\phi_l) \eea Now the
$d$-dimensional Hall conductance is given by \bea \sigma_d &=&
\sum_{\phi_1,\cdots,\phi_l=\phi_0,0}(-1)^{\sum_i\delta(\phi_i,0)}C(\phi_1,\cdots,\phi_l)  \nn \\
&=& C(\phi_0,\cdots,\phi_0,\phi_0) - C(\phi_0,\cdots,\phi_0,0) +
\cdots \nn \\ && -C(0,\cdots,0,0) \label{d-d} \eea where the delta
function satisfies $\delta(\phi_i,0)=1$ when $\phi_i = 0$, and
$\delta(\phi_i,0)=0$ when $\phi_i=\phi_0\equiv 2\pi$. When $d=4$(i.e.
$l=1$), Eq.(\ref{d-d}) reduces to Eq.(\ref{interacting}). The
original Niu-Thouless-Wu formula is also a special case of
Eq.(\ref{d-d}) with $d=2$(i.e. $l=0$).

\section{Fractional quantum Hall effects}\label{sec:fractional}

One of the main motivations for introducing the topological invariant
in Eq.(\ref{interacting}) is its potential applications in fractional
quantum Hall states. Before moving to higher dimensions, let us first
present a review of the Niu-Thouless-Wu formula of 2D fractional QH.
As has been known from the Ref.\cite{niu1985}, fractional
quantization of 2D Hall conductance is possible if the ground states
are degenerate on a 2D torus.

In 2D, the standard boundary condition is given\cite{niu1985} by
Eq.(\ref{condition-1}) except that the argument $\phi$ is absent.
Suppose that a 2D fractional quantum Hall system has $p$-fold
degenerate ground states
$|\Psi_1(\theta_1,\theta_2)\ra,\cdots,|\Psi_p(\theta_1,\theta_2)\ra$.
\footnote{It is worth noting that such a basis
$\{|\Psi_1\ra,\cdots,|\Psi_p\ra\}$ can be found only locally, namely
that they can be defined only in a topologically trivial patch on the
2D torus parameterized by $(\theta_1,\theta_2)$. The quantities we
actually need, such as ${\rm Tr} F_{12}(\theta_1,\theta_2)$, are
independent on the basis choices.} The Hall conductance is given by
an average over these degenerate ground states as\cite{niu1985}
(recall that we have taken the units $e=h=c=1$) \bea   \sigma_{2D}
&=& \frac{1}{p} \frac{1}{2\pi }\int_0^{2\pi} d\theta_1 d\theta_2
\sum_{\alpha=1}^p [\la \partial_{\theta_1} \Psi_\alpha|\partial_{\theta_2}\Psi_\alpha\ra -  \la \partial_{\theta_2}\Psi_\alpha|\partial_{\theta_1}\Psi_\alpha\ra ] \nn \\
&=& \frac{1}{p} \frac{1}{2\pi  }\int_0^{2\pi} d\theta_1 d\theta_2 {\rm Tr} F_{12}(\theta_1,\theta_2) \nn \\
&=& \overline{C}_1  \label{c1-fractional-2} \eea  where the matrix
elements of  the non-Abelian Berry curvature $F_{ij}$ read
$F_{ij}^{\alpha\beta}=\partial_i a_j^{\alpha\beta} -\partial_j
a_i^{\alpha\beta} +i[a_i,a_j]^{\alpha\beta}$, in which
$a_i^{\alpha\beta}=-i\la\Psi_\alpha(\theta_1,\theta_2 )|
\partial_{\theta_i} |\Psi_\beta(\theta_1,\theta_2 )\ra$ is the non-Abelian Berry
connection. The average Chern number $\overline{C}_1 \equiv
\frac{1}{p}C_1 \equiv \frac{1}{p }\frac{1}{2\pi }\int_0^{2\pi}
d\theta_1 d\theta_2 {\rm Tr} F_{12}(\theta_1,\theta_2)$, where $C_1$
is the standard definition of the first Chern
number\cite{nakahara1990} of the $U(p)$ fiber bundle. Note that the
$i[a_i,a_j]$ term in $F_{ij}$ vanishes after the tracing. It is a
mathematical fact that the first Chern number $C_1$ is quantized as
an integer, therefore, the Hall conductance is quantized as a
rational number with denominator $p$.

Eq.(\ref{c1-fractional-2}) can be rewritten as\cite{niu1985} \bea
\sigma_{2D} = \frac{1}{p} \frac{1}{2\pi i }\int_0^{2\pi p} d\theta_1
\int_0^{2\pi}d\theta_2 [\la \partial_{\theta_1}
\Psi_1|\partial_{\theta_2}\Psi_1\ra -  \la
\partial_{\theta_2}\Psi_1|\partial_{\theta_1}\Psi_1\ra ]
\label{c1-fractional} \eea where we have picked up a ground state
$\Psi_1$ from the degenerate ground state $\Psi_1,\cdots,\Psi_p$. The
parameter space has been enlarged to $(0<\theta_1 <2\pi p, 0<\theta_2
<2\pi)$.

Now let us move to higher dimensions. For a 4D fractional QH system,
suppose that the ground states are $p$-fold degenerate on the 4D
torus $T^4$ with boundary conditions described in Sec.\ref{sec:4d},
in other words, the ground states form a $U(p)$ bundle over the 2D
torus with coordinates $(\theta_1,\theta_2)$ ( Note that $\phi$ is
fixed ). \footnote{For simplicity, we assume that the degenerate
ground states cannot be divided into smaller subspaces, with each
subspace forming a fiber bundle over the torus parameterized by
$(\theta_1,\theta_2)$. Otherwise we can just pick up one subspace
that cannot be reduced further into smaller subspaces, and everything
discussed in this section will be unchanged. } We can define the
Berry connection $a_i^{\alpha\beta}(\theta_1,\theta_2,\phi)
=-i\la\Psi_\alpha(\theta_1,\theta_2,\phi )|
\partial_{\theta_i} |\Psi_\beta(\theta_1,\theta_2,\phi )\ra$ and the Berry
curvature $F_{ij}^{\alpha\beta}=\partial_i a_j^{\alpha\beta}
-\partial_j a_i^{\alpha\beta} +i[a_i,a_j]^{\alpha\beta}$.
Eq.(\ref{chern-4d}) can be straightforwardly generalized as \bea
C(\phi)=\frac{1}{2\pi} \int_0^{2\pi} d\theta_1 d\theta_2 {\rm Tr}
F_{12}(\theta_1,\theta_2,\phi) \label{chern-4d-fractional} \eea Note
that in Sec.\ref{sec:4d} we considered non-degenerate ground state,
therefore, the symbol ``${\rm Tr}$'' in
Eq.(\ref{chern-4d-fractional}) is absent Eq.(\ref{chern-4d}). We can
also define the  average  (first) Chern number for 4D QH as \bea
\overline{C}(\phi)=C(\phi)/p \eea By analogy with
Eq.(\ref{c1-fractional-2}), the 4D Hall conductance $\sigma_{4D}$ for
fractional quantum Hall effects is obtained as \bea \sigma_{4D}=
\overline{C}(\phi_0)-\overline{C}(0) \label{fractional-4d} \eea
Eq.(\ref{fractional-4d}) is among the central results of this paper.
In the presence of time reversal symmetry, the second term vanishes.
Eq.(\ref{fractional-4d}) reduces to Eq.(\ref{interacting}) when
$p=1$, namely the case without ground state degeneracy.

To conclude this section, we mention that the generalization of
Eq.(\ref{d-d}) for $d=2l+2$ dimensional fractional states read
$\sigma_d =
\sum_{\phi_1,\cdots,\phi_l=\phi_0,0}(-1)^{\sum_i\delta(\phi_i,0)}
\overline{C}(\phi_1,\cdots,\phi_l)$.

\section{More topological invariant for 4D fractional QH}\label{sec:more}

Having studied the 4D fractional Hall conductance using the
$(\theta_1,\theta_2,\phi)$ boundary conditions we have chosen, let us
investigate other choices of boundary conditions. The simplest choice
is \bea \Psi(\br_1,\cdots,\br_k +L_i\hat{\bx}_i,\cdots,\br_N;
\theta_1,\theta_2,\theta_3,\theta_4) \nn \\ = \exp(i\theta_i) \Psi
(\br_1,\cdots,\br_k,\cdots,\br_N;
\theta_1,\theta_2,\theta_3,\theta_4)  \label{standard} \eea for
$i=1,2,3,4$.  Suppose that the ground states are $p$-fold degenerate,
then these ground states form an $U(p)$ fiber bundle on the 4D torus
parameterized by $(\theta_1,\theta_2,\theta_3,\theta_4)$ with $0\leq
\theta_i<2\pi$. We can define a natural topological invariant \bea
C_2 =\frac{1}{32\pi^2}\int  d^4\theta \epsilon^{ijkl} {\rm
Tr}F_{ij}F_{kl} \label{chern-2} \eea where the matrix elements of
non-Abelian Berry curvature $F_{ij}$ are defined as
$F_{ij}^{\alpha\beta}(\theta_1,\theta_2,\theta_3,\theta_4)=\partial_i
a_j^{\alpha\beta} -\partial_j a_i^{\alpha\beta}
+i[a_i,a_j]^{\alpha\beta}$, where $i,j=1,2,3,4$. Eq.(\ref{chern-2})
is a second Chern number defined for fractional QH states in 4D. It
should be not be confused with the (lower-case) $c_2$ in
Eq.(\ref{noninteracting}), which is defined in terms of the free
Bloch states of noninteracting systems.

For $2n$-dimensional quantum Hall effects, we can straightforwardly
generalize $C_2$ to
$C_n$ as \bea C_n &=& \frac{1}{n!}\int {\rm Tr}(\frac{F}{2\pi})^n \nn \\
&=& \frac{1}{2^n n!(2\pi)^n}\int d^{2n}\theta
\epsilon^{\alpha_1\cdots\alpha_{2n}} {\rm Tr}
F_{\alpha_1\alpha_2}\cdots F_{\alpha_{2n-1}\alpha_{2n}}
\label{chern-n} \eea which are topological invariants for
higher-dimensional fractional QH states.

In 2D, the first Chern number $C_1$ of the $U(p)$ bundle is
proportional to the Hall conductance $\sigma_{2D}$. In fact,
Eq.(\ref{c1-fractional-2}) tells us that $C_1=p\sigma_{2D}$, thus
$C_1$ does not give us new topological invariant other than
$\sigma_{2D}$ and $p$. However, the 4D case is quite different. The
key difference between 2D and 4D is as follows. For the 2D QH, both
$\sigma_{2D}$ and $C_1$ are defined under the same boundary condition
parameterized by $(\theta_1,\theta_2)$. For 4D quantum Hall
insulators, the topological invariants $C_2$ and $\sigma_{4D}$ is
defined using different boundary conditions [ Eq.(\ref{condition-1}),
Eq.(\ref{condition-2}), and Eq.(\ref{condition-3}) for $\sigma_{4D}$,
but Eq.(\ref{standard}) for $C_2$ ], therefore, there is no direct
relation between $C_2$ and $\sigma_{4D}$. In principle, $C_2$ can
take different values given the same value of ground state degeneracy
$p$ and Hall coefficient $\sigma_{4D}$. The topological invariants
$C_2$ suggests that there are rich structures in 4D quantum Hall
effects. Higher dimensional QHs are similar: Higher Chern numbers
$C_n$($n=2,3,\cdots$) are not directly related to the Hall
coefficient $\sigma_{d}$ because they are defined under different
boundary conditions.

\section{Topological insulators in one-dimension}\label{sec:1d}

In this section we will briefly discuss 1D topological insulators to
prepare us for the investigation of 3D topological insulator in
Sec.\ref{sec:3d}. One-dimensional topological insulators can be
characterized by a $\theta$ term\cite{qi2008} \bea S_{{\rm
eff}}=\frac{1}{2\pi}\int dtdx \epsilon^{\mu\nu}\theta\partial_\mu
A_\nu \label{1d-theta} \eea Let us study the 1D insulator on a torus
$T^1$, which is just a circle. We take the boundary condition as \bea
\Psi(\br_1,\cdots,\br_k +L_1\hat{\bx}_1,\cdots,\br_N;  \theta_1) \nn
\\ = \exp(i\theta_1) \Psi (\br_1,\cdots,\br_k,\cdots,\br_N; \theta_1
) \eea namely that there is a gauge potential $A_1=\theta_1/L_1$.

Now there exists a simple topological invariant\cite{ortiz1994,guo2011}  \bea
\Gamma = \int_0^{2\pi} d\theta_1 a_1 (\theta_1)\label{1d-interacting} \eea
where the Berry connection  is defined as $a_1(\theta_1 ) =-i\la\Psi(
\theta_1 )|
\partial_{\theta_1} |\Psi (\theta_1 )\ra$. Eq.(\ref{1d-interacting}) is an
interacting generalization of the Zak phase\cite{zak1989}.
It has been applied to 1D
models\cite{guo2011,guo2012}, though its relation to $\theta$ term was not discussed.  Eq.(\ref{1d-interacting}) is defined modulo $2\pi$ because a local gauge transformation of the wavefunction can change it by $2\pi$.

When the ground state $|\Psi(\theta_1)\ra$ is not degenerate, the
$\theta$ value is given by $\theta =\Gamma$.  Since we are mainly
concerned with higher dimensional topological insulators, we will not
study applications of this 1D formula in details. It is useful to
mention that the quantity $\partial\theta/\partial\lambda$, where
$\lambda$ is a tuning parameter of the many-body Hamiltonian, is
usually more useful than $\theta$ itself, because
$\partial\theta/\partial\lambda$ does not have any ambiguity under
local gauge transformation of wavefunction\cite{ortiz1994}.

When the ground states are $p$-fold degenerate,   the natural
generalization of Eq.(\ref{1d-interacting}) is \bea \Gamma = \int
d\theta_1 {\rm Tr} a_1 (\theta_1)\label{1d-fractional} \eea where the
non-Abelian gauge potential is defined as
$a_1^{\alpha\beta}=-i\la\Psi_\alpha(\theta_1)
|\partial_{\theta_1}|\Psi_\beta(\theta_1)\ra$. The $\theta$ angle in
Eq.(\ref{1d-theta}) is given by \bea \theta=\overline{\Gamma}
 \label{1d-fractional} \eea where the average $\overline{\Gamma}$ is
defined as $\overline{\Gamma}=\frac{1}{p}\Gamma$.

\section{Topological insulators in three-dimensions: Integer and fractional}\label{sec:3d}

The approach we applied to 4D QH states can be naturally generalized
to 3D.  The 3D boundary conditions are chosen as follows. First, \bea
\Psi(\br_1,\cdots,\br_k +L_1\hat{\bx}_1,\cdots,\br_N; \theta_1, \phi)
\nn \\ = \exp(i\theta_1) \Psi (\br_1,\cdots,\br_k,\cdots,\br_N;
\theta_1, \phi) \label{condition-3d-1} \eea
 where $\br_k$ is the coordinate of the
$k$-th particle ( other variables such as spin are not shown for simplicity of notation ), and
$\hat{\bx}_1$ is the unit vector along the $x_1$ direction.   Second,
\bea \Psi (\br_1,\cdots,\br_k +L_2\hat{\bx}_2,\cdots,\br_N;
\theta_1,\phi) \nn
\\ = \exp(-i\phi\frac{ x_3}{L_3}) \Psi
(\br_1,\cdots,\br_k,\cdots,\br_N; \theta_1, \phi)
\label{condition-3d-2} \eea  and \bea \Psi (\br_1,\cdots,\br_k
+L_3\hat{\bx}_3,\cdots,\br_N; \theta_1, \phi) \nn \\ = \Psi
(\br_1,\cdots,\br_k,\cdots,\br_N; \theta_1, \phi)
\label{condition-3d-3} \eea where $\phi$ satisfies the same
quantization condition as discussed in Sec.\ref{sec:4d}. Now the
$\theta$ angle in Eq.(\ref{theta}) is proposed (for the cases without
ground state degeneracy) as \bea \theta=\Gamma(\phi_0)-\Gamma(0)
\label{3d} \eea where $\phi_0\equiv 2\pi$, and \bea
\Gamma(\phi)=\int_0^{2\pi} d\theta_1 a_1(\theta_1,\phi) \eea
$a_1(\theta_1,\phi) =-i\la\Psi( \theta_1,\phi)|
\partial_{\theta_1} |\Psi (\theta_1,\phi)\ra$
being the Berry connection defined in terms of the ground state
wavefunction. One can derive Eq.(\ref{3d}) by calculating the Berry
phase gained by the adiabatic evolution $A_1\rw A_1+2\pi/L_1$.   Due
to the topological terms $\frac{\theta}{4\pi^2}\partial_0 A_1
(\partial_2 A_3-\partial_3 A_2)$ contained in the $\theta$ term, when
a flux $\phi$ exists in $T_{23}$, as Eq.(\ref{condition-3d-2}) and
Eq.(\ref{condition-3d-3}) indicate,  the adiabatic evolution of
$A_1\rw A_1+2\pi/L_1$ generates a topological phase
$\theta\phi/2\pi$, which should be identified as the Berry phase
accumulated by the adiabatic evolution of ground state wavefunction,
namely $\int d\theta_1 a_1(\theta_1,\phi)$. It follows that
Eq.(\ref{3d}) is the formula for $\theta$. Note that potentially
there is another term $\theta'\partial_0 A_1$ that can contribute to
the Berry phase in the evolution $A_1\rw A_1+2\pi/L_1$, which is the
reason why the second term in Eq.(\ref{3d}) appears.

If the Hamiltonian and the ground state depend on a tuning parameter,
which we denote as $\theta_2$, then $\theta$ is a function of
$\theta_2$. The derivative of $\theta$ with respect to $\theta_2$ is
given by the gauge-invariant formula \bea
\frac{\partial\theta}{\partial \theta_2} = \int_0^{2\pi}d\theta_1
[F_{21}(\theta_1,\theta_2,\phi_0)-F_{21}(\theta_1,\theta_2, 0)] \eea
where $F_{21}(\theta_1,\theta_2,\phi
)=\partial_{\theta_2}a_1-\partial_{\theta_1}a_2$, and
$a_i(\theta_1,\theta_2,\phi )=-i\la\Psi (\theta_1,\theta_2,\phi )|
\frac{\partial}{\partial \theta_i} |\Psi (\theta_1,\theta_2,\phi
)\ra$. Similar to the 1D case discussed in Sec.\ref{sec:1d}, the
quantity $ \partial\theta/\partial \theta_2$ is usually more useful
than $\theta$ itself, because $ \partial\theta/\partial \theta_2$ is
invariant under any local gauge transformation of the wavefunction.

We will apply Eq.(\ref{3d}) to a noninteracting Dirac model in Appendix
\ref{sec:3d-dirac}, which gives the same result as
obtained\cite{qi2008} from Eq.(\ref{CS-3d}).

In the above calculations we have assumed that the $\theta$ term is
isotropic, which is always satisfied if there is time reversal
symmetry (though the Maxwell terms are generally still anisotropic).
If the $\theta$ term is anisotropic\cite{essin2010,malashevich2010},
namely that we have $\chi_{ij}E_i B_j=\chi_{ij}E_i \epsilon_{jkl}(\partial_k A_l -\partial_l A_k)$, we should calculate each
coefficient $\chi_{ij}$ separately, which is also given by Eq.(\ref{3d}) except that the twisted phase $\theta_1$ in Eq.(\ref{condition-3d-1}) is added in the $x_i$ direction instead of the $x_1$ direction, and the flux $\phi$ [ see Eq.(\ref{condition-3d-2}) and Eq.(\ref{condition-3d-3})] is added in the $(x_k,x_l)$ plane.

For 3D fractional states with $p$-fold ground state degeneracy, we
can generalize Eq.(\ref{3d}) as \bea
\theta=\overline{\Gamma}(\phi_0)-\overline{\Gamma}(0)
  \label{3d-fractional} \eea
where $\overline\Gamma(\phi)\equiv \frac{1}{p}\int_0^{2\pi}
d\theta_1{\rm Tr}a_1(\theta_1,\phi)$. The logic is similar to
Sec.\ref{sec:fractional}. An important feature is notable here. We
have the transformation rule $a_1\rw Ua_1 U^\dag +iU\partial U^\dag$
under a local gauge transformation of the basis of ground state
wavefunction, where $U=U(\theta_1,\phi)$ is a $p\times p$ unitary
matrix. This may change $\Gamma(\phi)$ by multiples of $2\pi/p$,
therefore, the $\theta$ angle of fractional topological insulators is
determined modulo $2\pi/p$.

As a digression, let us briefly mention the generalization for
$d=2l+1$ (spatial) dimensional (isotropic) $\theta$ term if the
system does not have ground state degeneracy on a $d$ dimensional
torus. The formula reads \bea \theta_d &=&
\sum_{\phi_1,\cdots,\phi_l=\phi_0,0}(-1)^{\sum_i\delta(\phi_i,0)}\Gamma(\phi_1,\cdots,\phi_l)  \nn \\
&=& \Gamma(\phi_0,\cdots,\phi_0,\phi_0) -
\Gamma(\phi_0,\cdots,\phi_0,0) + \cdots \nn \\ &&
-\Gamma(0,\cdots,0,0) \label{d-theta} \eea which is analogous to
Eq.(\ref{d-d}). The meanings of the arguments $\phi_1,\cdots,\phi_l$
are similar to that of Eq.(\ref{d-d}), which we shall not repeat
here. If the state is fractional, we have $\theta_d =
\sum_{\phi_1,\cdots,\phi_l=
\phi_0,0}(-1)^{\sum_i\delta(\phi_i,0)}\overline{\Gamma}(\phi_1,\cdots,\phi_l)$,
where $\overline{\Gamma}(\phi_1,\cdots,\phi_l)\equiv
\Gamma(\phi_1,\cdots,\phi_l)/p$, the integer $p$ being the ground
state degeneracy.

\section{Conclusions}\label{sec:conclusion}

In this paper we have defined precise topological invariants in terms
of the ground state wavefunctions on a torus.   This approach provides a conceptual framework in which many topological invariants and topological-field-theoretical
coefficients, such as $\sigma_{4D}$ (in 4D) and $\theta$ (in 3D),  acquire precise definitions even in the presence of arbitrary interaction and disorder.

Numerically, we do not expect that the wavefunction (on a torus)
approach followed in the present paper will be as efficient as the
topological Hamiltonian approach\cite{wang2012a,wang2013} mentioned
in Sec.\ref{sec:intro}. However, the present approach has a wider
range of validity because it is applicable in the presence of
arbitrary interaction and disorder, therefore, the present approach
is highly desirable for certain purposes, especially when both
interaction and disorder are present, or when the interaction is so
strong that exotic fractional states are generated. It is also useful
to note that the topological invariants in the present paper can also
be applied to bosonic topological insulators, for which other
topological invariants are hard to define.

\section{Acknowledgements}

ZW would like to thank  Liang Kong and Yong-Shi Wu for helpful
discussions.  ZW is supported by NSFC under Grant No. 11304175. SCZ
is supported by the Department of Energy, Office of Basic Energy
Sciences, Division of Materials Sciences and Engineering, under
contract DE-AC02-76SF00515.

\appendix

\section{Application in a three-dimensional noninteracting
model}\label{sec:3d-dirac}

In the noninteracting limit, Eq.(\ref{3d}) should give the same $\theta$ as the noninteracting formula\cite{qi2008}.
In this appendix we will check this in a simple noninteracting model. This appendix follows similar calculations of Sec.\ref{sec:4d-dirac}.

Let us study a simple 3D noninteracting Dirac model given as \bea
h(\bk)=v \sin k_1\tau^1+[v\sin k_2\sigma^1+v\sin k_3\sigma^2
+M(\bk)\sigma^3]\tau^3
 \eea where $M(\bk)=m+3-\sum_{i=1}^3\cos k_i$. In the
limit that $|m|<<1$, the low energy physics is dominated by the
$k\approx 0$ region, and we can linearly expand $h(\bk)$ as \bea
h(\bk)\approx vk_1\tau^1+(vk_2\sigma^1+vk_3\sigma^2+m\sigma^3)\tau^3
\label{linear-3d} \eea The boundary conditions are given in
Eq.(\ref{condition-3d-1}), Eq.(\ref{condition-3d-2}), and
Eq.(\ref{condition-3d-3}), which mean that there is a flux $\phi$
inside the 2D torus $T_{23}$. First let us take $\phi=\phi_0\equiv
2\pi$. By a calculation similar to Sec.\ref{sec:4d-dirac}, we can
first solve $h'(k_2,k_3)= vk_2\sigma^1+vk_3\sigma^2+m\sigma^3$ after
replacing $k_i\rw -i(\partial_i -A_i)$, whose eigenvalues read \bea
E_0=m; \,\,\, E_{n\pm} = \pm\sqrt{ m^2+2nBv^2}\, (n=1,2,\cdots)
\label{eigen-3d} \eea and the corresponding wavefunctions are
$(\psi_0,0)^T$ and $(\psi_n,\pm\psi_{n-1})^T$. Now we put these
eigenvalues back into the parenthesis of Eq.(\ref{linear-3d}), then
we have a serial of 1D Hamiltonians \bea h_0 &=&  v k_1\tau^1  +m\tau^3; \nn\\
h_{n\pm}&=& v k_1\tau^1  +E_{n\pm}\tau^3\,(n=1,2,\cdots ) \eea

Now the $\Gamma(\phi_0)$ in Eq.(\ref{3d}) can be found as
$\frac{\pi}{2}[{\rm sgn}(E_0)+\sum_{n }\sum_{\alpha=\pm} {\rm
sgn}(E_{n\alpha})] =\frac{\pi}{2}{\rm sgn}(m)$, which is similar to
Sec.\ref{sec:4d-dirac}. Again, due to the high-energy regularization,
we can only assert that $\Gamma(\phi_0)=\frac{\pi}{2}{\rm
sgn}(m)+{\rm constant}$. Consideration similar to
Sec.\ref{sec:4d-dirac} leads to $\Gamma(\phi_0)=\frac{\pi}{2}({\rm
sgn}(m)-1)$. Similarly we have $\Gamma(0)=0$, therefore, from
Eq.(\ref{3d}) it follows that
\bea \theta=\Gamma(\phi_0)-\Gamma(0)=\frac{\pi}{2}({\rm sgn}(m)-1) \eea
which means that $\theta=-\pi$ when $m<0$. This is consistent with
the result obtained from the noninteracting Chern-Simons
term\cite{qi2008}.

\bibliography{wave_function_4D}

\end{document}